\title{Equations of Motion of Schwarzschild, Reissner-Nordstr\"om and Kerr Particles}

\author{Peter A. Hogan\footnote{Email: peter.hogan@ucd.ie}, \\
        School of Physics\\
        University College Dublin\\
        Belfield, Dublin 4, Ireland}

\date{\today}

\documentclass[11pt,english]{article}
\usepackage[]{graphicx}

\begin{document}
\maketitle

\begin{abstract}
A technique for extracting from the appropriate field equations the relativistic motion of Schwarzschild, Reissner-Nordstr\"om and Kerr particles moving in
external fields is motivated and illustrated. The key assumptions are that (a) the particles are isolated and (b) near the particles the wave fronts of the radiation generated
by their motion are smoothly deformed spheres. No divergent integrals arise in this approach. The particles are not test particles. The formalism is used, however, to
derive the Mathisson--Papapetrou equations of motion of spinning test particles, neglecting spin-spin terms.
\end{abstract}

\section{Introduction}
We describe a method for modeling the relativistic motion of (uncharged or charged) particles moving in external gravitational and electromagnetic fields in general relativity. The particles modify the 
fields in which they move and thus are \emph{not} test particles. The modified fields near the particles are predominantly the Schwarschild, Reissner--Nordstr\"om or Kerr fields and thus 
the particles will be referred to as Schwarschild, Reissner--Nordstr\"om or Kerr particles respectively. The origin of the approach described here is the seminal and challenging paper 
by Robinson and Robinson \cite{RR:1972}, with the disconcertingly modest title of ``Equations of Motion in the Linear Approximation", published in the festschrift in honor of Professor J. L. Synge. This paper inspired early work involving space--times which are more special than those required here 
and which can be found in a series of papers by Hogan and Imaeda \cite{HI1}, \cite{HI2}, \cite{HI3} (see also \cite{HOB}). A line element, which is a byproduct of a study of gravitational radiation from bounded sources \cite{HT}, 
plays a key role in the development of our method. An early application can be found in \cite{HR}. The most up-to-date references to our technique and applications are in the paper \cite{FHI} and the 
books \cite{AFH} and \cite{BH}. Further development of this work has also involved graduate students Takashi Fukumoto and Shinpei Ogawa in 
Tohoku University, Aishling Nic an Tuile in University College Dublin and Florian Bolgar in 
\'Ecole Normale Sup\'erieure, Paris and has taken the form of M.Sc. and Ph.D. theses and an Internship Report.

To introduce our approach we begin with the Eddington--Finkelstein form of the Schwarzschild line element:
\begin{equation}\label{1}
ds^2=-r^2p_0^{-2}(dx^2+dy^2)+2\,du\,dr+\left (1-\frac{2\,m}{r}\right )du^2\ ,\end{equation}with 
\begin{equation}\label{2}
p_0=1+\frac{1}{4}(x^2+y^2)\ .\end{equation}Here $x, y$ are stereographic coordinates taking values in the ranges $-\infty<x<+\infty$ and $-\infty<y<+\infty$, $u$ is a null coordinate (in the 
sense that the hypersurfaces $u={\rm constant}$ are null) with $-\infty<u<+\infty$. The generators of these null hypersurfaces are null geodesics labelled by $(x, y)$ and $r$ (with $0\leq r<+\infty$) is 
an affine parameter along them. When the mass parameter $m=0$ the space--time is Minkowskian with line element
\begin{equation}\label{3}
ds_0^2=-r^2p_0^{-2}(dx^2+dy^2)+2\,du\,dr+du^2=\eta_{ij}dX^idX^j\ ,\end{equation}with $X^i=(X, Y, Z, T)$ rectangular Cartesian coordinates and time and $\eta_{ij}={\rm diag}(-1, -1, -1, +1)$. In this 
\emph{background space--time} with line element (\ref{3}) the hypersurfaces $u={\rm constant}$ are future null cones with vertices on the \emph{time--like geodesic} $r=0$. We can consider
\begin{equation}\label{4}
\gamma_{ij}\,dx^i\,dx^j=-\frac{2\,m}{r}du^2\ ,\end{equation}in coordinates $x^i=(x, y, r, u)$, as a perturbation of this background which is singular on the time--like geodesic $r=0$ and which, when added 
to the line element (\ref{3}) produces the line element (\ref{1}). Although we have no need to assume it now, we shall, in the sequel, assume that $m$ is small of first order and write $m=O_1$ 
in order to simplify the complicated calculations that develop. We shall consider the background space--time a model of the \emph{external field} in which the small mass is located. In the example here the 
background space--time is Minkowskian space--time and thus there is no external field present. Also in this example the time--like geodesic equations (satisfied by $r=0$ in (\ref{3})) are the \emph{equations 
of motion} of the small mass. 

In the space--time with line element (\ref{3}) let $X^i=w^i(u)$ be an arbitrary time--like world line with $u$ proper time or arc length along it. Then $v^i(u)=dw^i/du$ is the unit tangent to this line and satisfies 
$v_jv^j=\eta_{ij}v^iv^j=+1$ along the world line.  Equivalently $v^i$ is the 4--velocity of the particle with world line $X^i=w^i(u)$. Thus $a^i=dv^i/du$ is the 4--acceleration and satisfies $a_jv^j=\eta_{ij}a^iv^j=0$ along the 
world line. We shall lower and raise indices, referring to the coordinates $X^i$, using $\eta_{ij}$ and $\eta^{ij}$ respectively, with the latter defined by $\eta^{ij}\eta_{jk}=\delta^i_k$. 
We note that $v^i$ and $a^i$ are the components of the 4--velocity and 4--acceleration of the particle expressed in the coordinates $X^i$. These features of the world line can be 
incorporated in the Minkowskian line element by replacing (\ref{3}) by
\begin{equation}\label{5}
ds_0^2=-r^2P_0^{-2}(dx^2+dy^2)+2\,du\,dr+(1-2h_0r)du^2\ ,\end{equation}
with
\begin{equation}\label{6}
P_0=xv^1(u)+yv^2(u)+\left\{1-\frac{1}{4}(x^2+y^2)\right\}v^3(u)+\left\{1+\frac{1}{4}(x^2+y^2)\right\}v^4(u)\ ,\end{equation}and
\begin{equation}\label{7}
h_0=\frac{\partial}{\partial u}(\log P_0)=a_j\,k^j\ ,\end{equation}with
\begin{equation}\label{8}
k^j=P_0^{-1}\left [-x\delta^j_1-y\delta^j_2-\left\{1-\frac{1}{4}(x^2+y^2)\right\}\delta^j_3+\left\{1+\frac{1}{4}(x^2+y^2)\right\}\delta^j_4\right ]\ .\end{equation}
In (\ref{5}) $r=0$ is the world line $X^i=w^i(u)$ with the 4--velocity components appearing in (\ref{5}) via the function $P_0$ and the 4--acceleration appearing via $h_0$. The future 
null cones with vertices on this world line are the null hypersurfaces $u={\rm constant}$. The generators of these null cones are the null geodesic integral curves of the vector 
field $k^j$, which satisfies $k_jk^j=0$ and $k_jv^j=+1$. On each future null cone the generators are labeled by the stereographic coordinates $x, y$ and $r\geq 0$ is an affine parameter 
along them. To see how Einstein's vacuum field equations can lead to the equations of motion as we have defined them in the paragraph above we consider a space--time with line element
\begin{equation}\label{9}
ds^2=-r^2P_0^{-2}(dx^2+dy^2)+2\,du\,dr+\left (1-2h_0r-\frac{2\,m}{r}\right )du^2\ ,\end{equation}
which is a generalization of the Schwarzschild space--time with line element (\ref{1}). The Ricci tensor components in coordinates $x^i=(x, y, r, u)$ are now given exactly via
\begin{equation}\label{10}
R_{ij}dx^idx^j=\frac{6\,m\,h_0}{r^2}\,du^2\ .\end{equation}With $m\neq0$ we see that the vacuum field equations $R_{ij}=0$ require $a_jk^j=h_0=0$ and this should hold for all possible $k^j$. 
Hence we must have, as equations of motion, the time--like geodesic equations 
\begin{equation}\label{11}
a^j=0\ ,\end{equation}for the world line $r=0$ in the Minkowskian background space--time. This simple example illustrates a connection between the equations of motion and the field equations. 
In general we shall find, however, that the field equations by themselves will not be sufficient to determine the equations of motion. They will have to be supplemented by a smoothness condition 
on the wave fronts of the possible radiation produced by the motion of the particle. A singular point on a wave front translates in space--time into a singularity along a generator, extending to future 
null infinity, of the null hypersurface history of the wave front and thus violating the concept of an isolated source. A useful example with which to illustrate this is to consider the relativistic 
motion of a Reissner--Nordstr\"om particle moving in external electromagnetic and gravitational fields.

\section{Reissner-Nordstr\"om Particle}
We construct here a model of a Reissner--Nordstr\"om particle of small mass $m=O_1$ and small charge $e=O_1$ moving in external vacuum gravitational and electromagnetic fields. In this 
case the background space--time is a general solution of the vacuum Einstein--Maxwell field equations. We shall require a knowledge of this space--time, and the Maxwell field, in the neighborhood of an arbitrary time--like 
world line $r=0$ (say). The charged mass will then be introduced as a perturbation of this space--time, which is singular on this world line in the background space--time (in similar fashion to (\ref{4}) in the space--time 
with line element (\ref{3})), in such a way that the perturbed space--time is, for small values of $r$, predominantly the Reissner--Nordstr\"om space--time and Maxwell field. Since we have indicated in Section 1 that 
we shall make use of the field equations along with conditions on the wave fronts of possible radiation (electromagnetic and/or gravitational in the current example) produced by the motion of our particle, and 
since the histories of wave fronts are null hypersurfaces in space--time, we shall work in the background space--time and in the perturbed background space--time in a coordinate system based on 
a family of null hypersurfaces. Thus we begin by writing the line element of the background space--time as
\begin{equation}\label{12}
ds^2=-(\vartheta^1)^2-(\vartheta^2)^2+2\,\vartheta^3\,\vartheta^4=g_{ab}\vartheta^a\,\vartheta^b\ ,\end{equation}with the 1--forms $\vartheta^1, \vartheta^2, \vartheta^3$ and $\vartheta^4$ given by
\begin{eqnarray}
\vartheta^1&=&r\,p^{-1}(e^\alpha\cosh\beta\,dx+e^{-\alpha}\sinh\beta\,dy+a\,du)\ ,\label{13}\\
\vartheta^2&=&r\,p^{-1}(e^\alpha\sinh\beta\,dx+e^{-\alpha}\cosh\beta\,dy+b\,du)\ ,\label{14}\\
\vartheta^3&=&dr+\frac{c}{2}\,du\ ,\label{15}\\
\vartheta^4&=&du\ .\label{16}\end{eqnarray}The constants $g_{ab}=g_{ba}$ constitute the components of the metric tensor on the tetrad defined by the basis 1--forms. 
Tetrad indices will be denoted by the early letters of the alphabet $a, b, c,\dots\ \ $, and tetrad indices will be lowered with $g_{ab}$ and raised 
with $g^{ab}$ where $g^{ab}g_{bc}=\delta^a_c$. This line element is completely general in that it involves six functions $p, \alpha, \beta, a, b, c$ each of the four coordinates $x, y, r, u$. It is thus equivalent to 
line elements constructed by Sachs \cite{Sachs} and by Newman and Unti \cite{NU}. It was constructed in \cite{HT} for the study of gravitational radiation from bounded sources and was 
specifically designed to have the Robinson--Trautman (\cite{RT1}, \cite{RT2}) line elements appear as a convenient special case (got by simply putting $\alpha=\beta=0$). This latter property 
also makes it convenient in the current context. We shall take $r=0$ to be the equation of an arbitrary time--like world line in the space--time with line element (\ref{12}). The space--time in the 
neighborhood of this world line has the well--known Fermi property (see, for example, \cite{LOR1}, \cite{LOR2}) that there exist coordinates in terms of which the metric tensor components satisfy
\begin{equation}\label{17}
g_{ij}=\eta_{ij}+O(r^2)\ .\end{equation}In view of the line element (\ref{5}) we can implement this property in the context of the line element (\ref{12}) by expanding the six functions appearing there in powers 
of $r$ as follows:
\begin{eqnarray}
p&=&P_0(1+q_2r^2+q_3r^3+\dots\ \ )\ ,\label{18}\\
\alpha&=&\alpha_2\,r^2+\alpha_3\,r^3+\dots\ \ \ ,\label{19}\\
\beta&=&\beta_2\,r^2+\beta_3\,r^3+\dots\ \ \ ,\label{20}\\
a&=&a_1\,r+a_2\,r^2+\dots\ \ \ ,\label{21}\\
b&=&b_1\,r+b_2\,r^2+\dots\ \ \ ,\label{22}\\
c&=&c_0+c_1\,r+c_2\,r^2+\dots\ \ \ ,\label{23}\end{eqnarray}where the coefficients of the powers of $r$ are functions of $x, y, u$. Following (\ref{5}) the function $P_0$ here is given 
by (\ref{6}) and 
\begin{equation}\label{24}
c_0=1=\Delta\log P_0\ \ \ {\rm with}\ \ \ \Delta=P_0^2\left (\frac{\partial^2}{\partial x^2}+\frac{\partial^2}{\partial y^2}\right )\ ,\end{equation}and
\begin{equation}\label{25}
c_1=-2h_0\ ,\end{equation}with $h_0$ given by (\ref{7}). The operator $\Delta$ is the Laplacian on the unit 2--sphere. The relationship between the rectangular Cartesian coordinates and time $X^i$ 
and the coordinates $x, y, r, u$ near the arbitrary time--like world line $r=0$ ($\Leftrightarrow X^i=w^i(u)$, introduced in Section 1) is given by (see, for example, \cite{NU2})
\begin{equation}\label{26}
X^i=w^i(u)+r\,k^i\ ,\end{equation}neglecting $O(r^2)$--terms.

At this point in the discussion we should note some formulas associated with Minkowskian space--time with line element (\ref{5}) or, equivalently, with metric tensor components $\eta_{ij}$ in 
coordinates $X^i$. A useful basis of 1--forms suggested by (\ref{5}) is
\begin{equation}
\omega^1=r\,P_0^{-1}dx\ ,\ \omega^2=r\,P_0^{-1}dy\ ,\ \omega^3=dr+\left (\frac{1}{2}-h_0\,r\right )du\ ,\ \omega^4=du\ .\label{27}\end{equation}When expressed in terms of the coordinates $X^i$ 
using the transformation (\ref{26}) these read
\begin{eqnarray}
\omega^1&=&-P_0\,\frac{\partial k_i}{\partial x}\,dX^i=-\omega_1\ ,\label{28}\\
\omega^2&=&-P_0\,\frac{\partial k_i}{\partial y}\,dX^i=-\omega_2\ ,\label{29}\\
\omega^3&=&\left (v_i-\frac{1}{2}k_i\right )\,dX^i=\omega_4\ ,\label{30}\\
\omega^4&=&k_i\,dX^i=\omega_3\ .\label{31}\end{eqnarray} In coordinates $X^i$ the vector fields with components $v^i, k^i, \partial k^i/\partial x, \partial k^i/\partial y$ form a basis and it 
is useful to express the metric tensor components $\eta^{ij}$ on this basis as
\begin{equation}\label{32}
\eta^{ij}=-P_0^2\left (\frac{\partial k^i}{\partial x}\frac{\partial k^j}{\partial x}+\frac{\partial k^i}{\partial y}\frac{\partial k^j}{\partial y}\right )+k^iv^j+k^jv^i-k^ik^j\ .\end{equation}Also the second 
partial derivatives of $k^i$ play a key role in the calculations and when they are expressed on this basis we have
\begin{eqnarray}
\frac{\partial^2k^i}{\partial x^2}&=&P_0^{-2}(v^i-k^i)-\frac{\partial}{\partial x}(\log P_0)\,\frac{\partial k^i}{\partial x}+\frac{\partial}{\partial y}(\log P_0)\,\frac{\partial k^i}{\partial y}\ ,\label{33}\\
\frac{\partial^2k^i}{\partial y^2}&=&P_0^{-2}(v^i-k^i)+\frac{\partial}{\partial x}(\log P_0)\,\frac{\partial k^i}{\partial x}-\frac{\partial}{\partial y}(\log P_0)\,\frac{\partial k^i}{\partial y}\ ,\label{34}\\
\frac{\partial^2k^i}{\partial x\partial y}&=&-\frac{\partial}{\partial y}(\log P_0)\,\frac{\partial k^i}{\partial x}-\frac{\partial}{\partial x}(\log P_0)\,\frac{\partial k^i}{\partial y}\ .\label{35}\end{eqnarray}
As a simple indication of the usefulness of these formulas we notice that if we add (\ref{33}) and (\ref{34}) we obtain
\begin{equation}\label{36}
\Delta k^i+2\,k^i=2\,v^i\ ,\end{equation}with the operator $\Delta$ defined in (\ref{24}). Taking the scalar product, with respect to the Minkowskian metric tensor $\eta_{ij}$, of this equation 
with the 4--acceleration $a^i$ yields
\begin{equation}\label{37}
\Delta h_0+2\,h_0=0\ ,\end{equation}with $h_0$ given by (\ref{7}). This demonstrates that $h_0$ is an $l=1$ spherical harmonic. A spherical harmonic $Q$ of order $l$ is a smooth 
solution, for $-\infty<x, y<+\infty$, of the equation $\Delta Q+l(l+1)Q=0$ for $l=0, 1, 2, \dots\ \ $.

Returning now to the construction of the background Einstein--Maxwell field, the potential 1--form leading to the background electromagnetic field is given by 
\begin{equation}\label{38}
A=L\,dx+M\,dy+K\,du\ ,\end{equation}where $L, M, K$ are functions of $x, y, r, u$. We have used the freedom to add an exact differential (i.e. a gauge transformation) to remove one 
component of this 1--form for convenience. The functions appearing in (\ref{38}) will be expanded in powers of $r$ in such a way that the resulting Maxwell 2--form $F=dA$, the exterior 
derivative of (\ref{38}), is non--singular at $r=0$, since this will be the \emph{external} Maxwell field. The simplest expansions to achieve this are:
\begin{eqnarray}
L&=&r^2\,L_2+r^3\,L_3+\dots\ \ \ \ ,\label{39}\\
M&=&r^2\,M_2+r^3\,M_3+\dots\ \ \ \ ,\label{40}\\
K&=&r\,K_1+r^2\,K_2+\dots\ \ \ \ ,\label{41}\end{eqnarray}with the coefficients of the powers of $r$ functions of $x, y, u$. The Maxwell 2--form is given by 
\begin{equation}\label{42}
F=dA=\frac{1}{2}F_{ab}\,\vartheta^a\wedge\vartheta^b\ ,\end{equation}where $F_{ab}=-F_{ba}$ are the components of the 2--form on the tetrad defined by the basis 1--forms (\ref{13})--(\ref{16}). 
The leading terms, in powers of $r$, of the tetrad components $F_{ab}$ which will interest us are, in light of the expansions (\ref{18})--(\ref{23}) and (\ref{39})--(\ref{41}),
\begin{equation}\label{43}
F_{13}=-2P_0\,L_2+O(r)\ ,\ \ F_{23}=-2P_0\,M_2+O(r)\ ,\ \ F_{34}=K_1+O(r)\ .\end{equation}Thus on $r=0$ we can replace the basis (\ref{13})--(\ref{16}) by the Minkowskian counterparts (\ref{28})--
(\ref{31}) and the coordinates $x, y, r, u$ by the coordinates $X^i$ following (\ref{26}) to arrive at
\begin{equation}\label{44}
L_2=\frac{1}{2}F_{ij}(u)\,k^i\,\frac{\partial k^j}{\partial x}\ ,\ M_2=\frac{1}{2}F_{ij}(u)\,k^i\,\frac{\partial k^j}{\partial y}\ ,\ \ K_1=F_{ij}(u)\,k^i\,v^j\ ,\end{equation}where $F_{ij}(u)=-F_{ji}(u)$ are 
the components of the external Maxwell tensor on the world line $r=0$ calculated in the coordinates $X^i$. With $k^i$ given by (\ref{8}) above we see that now the dependence of the 
functions $L_2, M_2, K_1$ on $x, y$ is explicitly known. We can therefore use the useful formulas (\ref{33})--(\ref{35}) to verify that Maxwell's vacuum field equations, 
$d{}^*F=0$, where the star denotes the Hodge dual of the 2--form $F$, are satisfied by $L_2, M_2, K_1$. The relevant equations obtained from the leading terms, in powers of $r$, in Maxwell's equations are
\begin{equation}\label{45}
K_1=P_0^2\left (\frac{\partial L_2}{\partial x}+\frac{\partial M_2}{\partial y}\right )\ ,\ \ \Delta K_1+2P_0^2\left (\frac{\partial L_2}{\partial x}+\frac{\partial M_2}{\partial y}\right )=0\ ,\end{equation}
and
\begin{equation}\label{46}
\frac{\partial K_1}{\partial x}+2\,L_2-\frac{\partial}{\partial y}\left\{P_0^2\left (\frac{\partial M_2}{\partial x}-\frac{\partial L_2}{\partial y}\right )\right\}=0\ ,\end{equation}
\begin{equation}\label{47}
\frac{\partial K_1}{\partial y}+2\,M_2+\frac{\partial}{\partial x}\left\{P_0^2\left (\frac{\partial M_2}{\partial x}-\frac{\partial L_2}{\partial y}\right )\right\}=0\ .\end{equation}We see that the second 
equation in (\ref{45}) is a consequence of (\ref{46}) and (\ref{47}) and the two equations in (\ref{45}) imply that $K_1$ is an $l=1$ spherical harmonic since
\begin{equation}\label{48}
\Delta K_1+2\,K_1=0\ .\end{equation}This latter equation can, of course, be verified directly using the last of (\ref{44}) and the formulas (\ref{33})--(\ref{35}).

The analogue, for the gravitational field, of the Maxwell 2--form with tetrad components $F_{ab}$ given by (\ref{43}) is the Weyl conformal curvature tensor with tetrad components $C_{abcd}$. The 
components which will be of interest to us in the neighborhood of the world line $r=0$ are given by
\begin{equation}\label{49}
C_{1313}+iC_{1323}=6\,(\alpha_2+i\beta_2)+O(r)\ ,\end{equation}and
\begin{equation}\label{50}
C_{3431}+iC_{3432}=\frac{3}{2}P_0^{-1}\left\{a_1+ib_1+4P_0^2\,\frac{\partial q_2}{\partial\bar\zeta}\right\}+O(r)\ ,\end{equation}where $\zeta=x+iy$ and a bar denotes complex conjugation. These 
must be examined in conjunction with the Einstein--Maxwell vacuum field equations for the background space--time:
\begin{equation}\label{51}
R_{ab}=2\,\left\{F_{ca}F^c{}_b-\frac{1}{4}g_{ab}F_{cd}F^{cd}\right\}=2\,E_{ab}\ ,\end{equation}where $R_{ab}$ are the tetrad components of the Ricci tensor of the background space--time, $g_{ab}$ are 
the tetrad components of the metric tensor and, as indicated following (\ref{16}), tetrad indices are raised using the components $g^{ab}$ of the inverse of the matrix with components $g_{ab}$ and 
$E_{ab}=E_{ba}$ are the tetrad components of the electromagnetic energy--momentum tensor. To satisfy the field equation $R_{33}=2E_{33}+O(r)$ we find that
\begin{equation}\label{52}
q_2=\frac{2}{3}\,P_0^2\left (L_2^2+M_2^2\right )\ .\end{equation}This can be rewritten, using $L_2$ and $M_2$ given by (\ref{44}) and the expression (\ref{32}) for the components 
of the inverse of the Minkowskian metric tensor, simply as
\begin{equation}\label{53}
q_2=-\frac{1}{6}F^p{}_i(u)\,F_{pj}(u)\,k^i\,k^j\ .\end{equation}Here $F_{pj}(u)$ are the components of the Maxwell tensor on the world line $r=0$ in coordinates $X^i$ while 
$F^p{}_i(u)=\eta^{pj}F_{ji}(u)$. Using this expression for $q_2$ in (\ref{50}) and the basis (\ref{28})--(\ref{31}), in similar fashion to the passage from (\ref{43}) to (\ref{44}) above, we conclude 
from (\ref{49}) and (\ref{50}) that
\begin{equation}\label{54}
\alpha_2=\frac{1}{6}P_0^2\,C_{ijkl}(u)\,k^i\,\frac{\partial k^j}{\partial x}\,k^k\,\frac{\partial k^l}{\partial x}\ ,\ \ \beta_2=\frac{1}{6}P_0^2\,C_{ijkl}(u)\,k^i\,\frac{\partial k^j}{\partial x}\,k^k\,\frac{\partial k^l}{\partial y}\ ,\end{equation}
and
\begin{eqnarray}
a_1&=&\frac{2}{3}P_0^2\,\left (C_{ijkl}(u)\,k^i\,v^j\,k^k\,\frac{\partial k^l}{\partial x}+F^p{}_i(u)\,F_{pj}(u)\,k^i\,\frac{\partial k^j}{\partial x}\right )\ ,\label{55}\\
b_1&=&\frac{2}{3}P_0^2\,\left (C_{ijkl}(u)\,k^i\,v^j\,k^k\,\frac{\partial k^l}{\partial y}+F^p{}_i(u)\,F_{pj}(u)\,k^i\,\frac{\partial k^j}{\partial y}\right )\ ,\label{56}\end{eqnarray}
where $C_{ijkl}(u)$ are the components of the Weyl tensor of the background space--time calculated on the world line $r=0$ in the coordinates $X^i$. With $P_0$ given 
by (\ref{6}) and $k^j$ by (\ref{8}), we see that the dependence of $q_2$ in (\ref{53}), $\alpha_2$ and $\beta_2$ in (\ref{54}), and $a_1$ and $b_1$ in (\ref{55}) and (\ref{56}) on the 
coordinates $x, y$ is explicitly known. The Einstein--Maxwell vacuum field equations $R_{12}=2E_{12}+O(r)$ and $R_{11}-R_{22}=2(E_{11}-E_{22})+O(r)$ yield the pair 
of real field equations incorporated in the single complex equation:
\begin{equation}\label{57'}
2(\alpha_2+i\beta_2)=-\frac{\partial}{\partial\bar\zeta}\left\{a_1+ib_1+4\,P_0^2\,\frac{\partial q_2}{\partial\bar\zeta}\right\}\ .\end{equation}The pair of real field equations $R_{13}=2E_{13}+O(r)$ 
and $R_{23}=2E_{23}+O(r)$ can be written as the complex equation:
\begin{equation}\label{58'}
a_1+ib_1+4\,P_0^2\,\frac{\partial q_2}{\partial\bar\zeta}=2\,P_0^4\,\frac{\partial}{\partial\zeta}\{P_0^{-2}(\alpha_2+i\beta_2)\}\ .\end{equation}The field equations (\ref{57'}) and (\ref{58'}) must be satisfied 
by $q_2, \alpha_2, \beta_2, a_1, b_1$ given by (\ref{53})--(\ref{56}). This important check can be carried out using the useful formulas (\ref{33})--(\ref{35}). We are now ready to introduce the Reissner--Nordstr\"om 
particle as a perturbation of this background space--time and Maxwell field.

The line element of the perturbed space--time is given by
\begin{equation}\label{57}
ds^2=-(\hat\vartheta^1)^2-(\hat\vartheta^2)^2+2\,\hat\vartheta^3\,\hat\vartheta^4=g_{ab}\hat\vartheta^a\,\hat\vartheta^b\ ,\end{equation}with the 1--forms $\hat\vartheta^1, \hat\vartheta^2, \hat\vartheta^3$ 
and $\hat\vartheta^4$ given by
\begin{eqnarray}
\hat\vartheta^1&=&r\,\hat p^{-1}(e^{\hat\alpha}\cosh\hat\beta\,dx+e^{-\hat\alpha}\sinh\hat\beta\,dy+\hat a\,du)\ ,\label{58}\\
\hat\vartheta^2&=&r\,\hat p^{-1}(e^{\hat\alpha}\sinh\hat\beta\,dx+e^{-\hat\alpha}\cosh\hat\beta\,dy+\hat b\,du)\ ,\label{59}\\
\hat\vartheta^3&=&dr+\frac{\hat c}{2}\,du\ ,\label{60'}\\
\hat\vartheta^4&=&du\ .\label{60}\end{eqnarray}To achieve our aim of having a predominantly Reissner--Nordstr\"om field for small values of $r$ we take the functions appearing here to have the 
following expansions in powers of $r$:
\begin{eqnarray}
\hat p&=&\hat P_0\,(1+\hat q_2\,r^2+\hat q_3\,r^3+\dots\ \ \ )\ ,\label{61}\\
\hat\alpha&=&\hat\alpha_2\,r^2+\hat\alpha_3\,r^3+\dots\ \ \ \ ,\label{62}\\
\hat\beta&=&\hat\beta_2\,r^2+\hat\beta_3\,r^3+\dots\ \ \ \ ,\label{63}\\
\hat a&=&\frac{\hat a_{-1}}{r}+\hat a_0+\hat a_1\,r+\hat a_2\,r^2+\dots\ \ \ \ ,\label{64}\\
\hat b&=&\frac{\hat b_{-1}}{r}+\hat b_0+\hat b_1\,r+\hat b_2\,r^2+\dots\ \ \ \ ,\label{65}\\
\hat c&=&\frac{e^2}{r^2}-\frac{2\,(m+2\,\hat f_{-1})}{r}+\hat c_0+\hat c_1\,r+\hat c_2\,r^2+\dots\ \ \ \ .\label{66}\end{eqnarray}
The coefficients of the various powers of $r$ here are functions of $x, y, u$. The hatted functions differ from their background values by $O_1$--terms. Thus in particular $\hat a_{-1}=O_1, \hat a_0=O_1, \hat b_{-1}=O_1, 
\hat b_0=O_1, $ but $\hat f_{-1}=O_2$. The perturbed potential 1--form 
\begin{equation}\label{67}
A=\hat L\,dx+\hat M\,dy+\hat K\,du\ ,\end{equation}is predominantly the Li\'enard--Wiechert 1--form ($=e\,(r^{-1}-h_0)\,du$) up to a gauge term so that
\begin{eqnarray}
\hat L&=&r^2\,\hat L_2+r^3\,\hat L_3+\dots\ \ \ \ ,\label{68}\\
\hat M&=&r^2\,\hat M_2+r^3\,\hat M_3+\dots\ \ \ \,\label{69}\\
\hat K&=&\frac{(e+\hat K_{-1})}{r}-e\,h_0+r\,\hat K_1+r^2\,\hat K_2+\dots\ \ \ \ ,\label{70}\end{eqnarray}where again the coefficients of the powers of $r$ here are functions of $x, y, u$ which differ from 
their background values by $O_1$--terms and $\hat K_{-1}=O_2$. The expansions (\ref{61})--(\ref{66}) and (\ref{68})--(\ref{70}) appear to us to be the minimal assumptions necessary to have a small charged 
mass with gravitational and electromagnetic fields predominantly those of a Reissner--Nordstr\"om particle for small values of $r$. They certainly restrict the model of such a particle moving in external 
gravitational and electromagnetic fields and their generalization or otherwise is a topic for further study.

When the metric tensor given by (\ref{57})--(\ref{60}) and the potential 1--form (\ref{67}), together with the expansions (\ref{61})--(\ref{66}) and (\ref{68})--(\ref{70}), are substituted into Maxwell's vacuum field equations and the Einstein--Maxwell vacuum field equations, and all terms are gathered on the 
left hand side of each equation with zero on the right hand side, we find a finite number of terms involving inverse powers of $r$ and an infinite number of terms involving positive (or zero) powers of $r$. In 
an ideal world we would equate the coefficients of each of these powers to zero and from the resulting equations derive the coefficients in the expansions (\ref{61})--(\ref{66}) and (\ref{68})--(\ref{70}). However the reality is 
that we can at best make the coefficients small in terms of $m$ and $e$. How small depends upon how accurately we require the equations of motion. Also the number of coefficients we require in the 
expansions (\ref{61})--(\ref{66}) and (\ref{68})--(\ref{70}) depends 
upon how accurately we require the equations of motion. For the purpose of the present illustration we shall calculate the equations of motion in first approximation and thus with an $O_2$--error. The 
null hypersurfaces $u={\rm constant}$ in the space--time with line element (\ref{57}) are the histories of possible wave fronts of radiation (electromagnetic and/or gravitational) produced by the 
motion of the Reissner--Nordstr\"om particle. For small values of $r$ the degenerate metric on these hypersurfaces is given by the line element
\begin{equation}\label{71}
ds_0^2=-r^2\hat P_0^{-2}(dx^2+dy^2)\ ,\end{equation}where $\hat P_0$ differs from its background value $P_0$ given in (\ref{6}) by $O_1$--terms and thus can be written
\begin{equation}\label{72}
\hat P_0=P_0\,(1+Q_1+O_2)\ ,\end{equation}with $Q_1(x, y, u)=O_1$. We will assume that the line elements (\ref{71}) are smooth deformations of the line element of a 2--sphere. This means that 
$Q_1$ is a well behaved function for $-\infty<x, y<+\infty$ and thus free of singularities in $x, y$. Now the field equations yield
\begin{eqnarray}
\hat a_{-1}&=&-4\,e\,P_0^2L_2+O_2=O_2\ ,\label{73}\\
\hat b_{-1}&=&-4\,e\,P_0^2M_2+O_2=O_1\ ,\label{74}\\
\hat c_0&=&1+\Delta Q_1+2\,Q_1+8\,e\,F_{ij}\,k^i\,v^j+O_2\ ,\label{75}\end{eqnarray}and
\begin{equation}\label{76}
-\frac{1}{2}\Delta(\Delta Q_1+2\,Q_1)=6\,m\,a_i\,k^i-6\,e\,F_{ij}\,k^i\,v^j+O_2\ .\end{equation}Both terms on the right hand side of (\ref{76}) are $l=1$ spherical harmonics and thus 
(\ref{76}) can be easily integrated (discarding the solution of the $l=0$ spherical harmonic equation which is singular for infinite values of $x$ and $y$) to read
\begin{equation}\label{77}
\Delta Q_1+2\,Q_1=6\,m\,a_i\,k^i-6\,e\,F_{ij}\,k^i\,v^j+A(u)+O_2\ ,\end{equation}where $A(u)=O_1$ is arbitrary (an $l=0$ spherical harmonic). We note that spherical harmonics corresponding 
to $l=0$ or $l=1$ in the perturbation $Q_1$ in (\ref{72}) are trivial, in the sense that the 2--sphere remains a 2--sphere under these perturbations, and can be neglected. Since the first two terms 
in (\ref{77}) are both $l=1$ spherical harmonics the solution $Q_1$ will necessarily have a singularity in $x, y$ (a \emph{directional singularity}) unless they combine to vanish or be at most small 
of second order. Thus we must have
\begin{equation}\label{78}
m\,a_i\,k^i=e\,F_{ij}\,k^i\,v^j+O_2\ ,\end{equation}for all possible values of $k^i$ and thus 
\begin{equation}\label{79}
m\,a_i=e\,F_{ij}\,v^j+O_2\ .\end{equation}We have arrived at the Lorentz equations of motion in first approximation.

The example above serves to illustrate our approach. If it is continued to the next order of approximation the equations of motion which emerge are \cite{FHI}
\begin{equation}\label{80}
m\,a_i=e\,F_{ij}\,v^j+\frac{4}{3}e^2\,h^k_i\,F^p{}_k\,F_{pj}\,v^j+\frac{2}{3}e^2\,h^k_i\,\dot a^k+e^2\,T_i+O_3\ ,\end{equation}where $h^i_j=\delta ^i_j-v^i\,v_j$ is the projection 
tensor and the dot denotes differentiation with respect to proper time $u$. The second term on the right hand side here is an $O_2$--correction to the external 4--force. The 
third term on the right hand side is the electromagnetic radiation reaction 4--force. The fourth term on the right hand side is a \emph{tail term} given by an integral with 
respect to proper time $u$ from $-\infty$ to the current proper time $u$ of a vector whose components are proportional to the external Maxwell field $F_{ij}(u)$ and involve functions of $u$ 
of integration and a pair of space--like vectors, which occur naturally, defined along the world line $r=0$ in the background space--time. When $e=0$ we obtain the equations of motion 
of a Schwarzschild particle in second approximation given by the approximate time--like geodesic equations 
\begin{equation}\label{81}
m\,a_i=O_3\ .\end{equation}
There is a formal similarity between the equations of motion (\ref{80}) and equations of motion derived by DeWitt and Brehme \cite{DeB}. However DeWitt and Brehme 
have removed an infinite term from their equations of motion and have considered a scenario (the world line of a charged test particle in a curved space--time, and hence utilize only Maxwell's 
equations on a curved space--time) which is different to that constructed here.

Finally, so that the reader is left in no doubt as to the procedure involved in the passage from (\ref{77}) to (\ref{79}), we can give a simplified example of the procedure. If 
$\lambda =\cos\theta$ for $0\leq\theta\leq\pi$ (or equivalently for $-1\leq\lambda\leq+1$) consider a function $f=f(\lambda)$ satisfying the inhomogeneous $l=1$ Legendre 
equation (i.e. the inhomogeneous $l=1$ spherical harmonic equation satisfied by a function of the single variable $\lambda$) with an $l=1$ Legendre polynomial on the right hand side:
\begin{equation}\label{82}
\Delta f+2\,f\equiv\frac{d}{d\lambda}\left\{(1-\lambda^2)\,\frac{df}{d\lambda}\right\}+2\,f=t_1\,P_1(\lambda)\ ,\end{equation}where $t_1$ is a constant. For this equation to have a solution 
which is non--singular at $\lambda=\pm1$ (which are directional singularities) we must have $t_1=0$ (analogous to (\ref{78})). This easily follows from consideration of the general solution
\begin{equation}\label{83}
f(\lambda)=-\frac{1}{6}\,t_1\,\lambda\,\log(1-\lambda^2)+t_2\,P_1(\lambda)+t_3\,Q_1(\lambda)\ ,\end{equation}where $t_2, t_3$ are constants and $Q_1(\lambda)$ is the Legendre function corresponding 
to $l=1$. Of course the required directional singularity--free solution also requires us to take $t_3=0$ here.

\section{Kerr Particle}
We outline here the application of our technique to the construction of a model of a spinning particle, or Kerr particle, moving in an external vacuum gravitational field, and the derivation of its equations of motion. This work, 
carried out by Shinpei Ogawa and the author, is reported in \cite{BH}. Although we consider only the equations of motion in first approximation here, the calculations are more intricate 
than those necessary for the study of the Reissner--Nordstr\"om particle. Consequently we must refer the interested reader to \cite{BH} for further details.

The fundamental aspects of our technique are the construction of a background space--time with an arbitrary time--like world line in the neighborhood of which the space-time is flat, following from 
the Fermi property, and then the introduction of the particle of interest as a perturbation of this space--time. In the flat neighborhood of the world line in the background space--time we wrote the 
line element in the form (\ref{5}) which introduces the 4--velocity and 4--acceleration of the world line into the line element. To consider a particle with spin in this context we introduce the spin variables 
(what will become the three independent components of the angular momentum per unit mass of the particle) into the line element of Minkowskian space--time to accompany the 4--velocity and 
4--acceleration already present. We do this in such a way that if $a^i=0$ then the form of the line--element of Minkowskian space--time coincides with the Kerr line element, with three components 
of angular momentum per unit mass, in the special case in which the mass $m=0$. The latter form of the Kerr line element can be found in (\cite{BarHogbook}, p.37). Let $s^i(u)$ be the components of 
the \emph{spin vector} in coordinates $X^i$, defined along the arbitrary time--like world line $X^i=w^i(u)$ to be everywhere orthogonal to the 4--velocity $v^i(u)$ and Fermi transported, thus, 
\begin{equation}\label{84}
s^i\,v_i=0\ \ \ {\rm and}\ \ \ \frac{ds^i}{du}=-(a_j\,s^j)\,s^i\ ,\end{equation}respectively.  The equivalent \emph{spin tensor} is defined by
\begin{equation}\label{85}
s_{ij}=\epsilon_{ijkl}\,v^k\,s^l=-s_{ji}\ ,\end{equation}where $\epsilon_{ijkl}$ are the components of the Levi--Civit\`a permutation tensor in coordinates $X^i$. We introduce $s^i$ into the Minkowskian line element 
by replacing (\ref{26}) by
\begin{equation}\label{86}
X^i=w^i(u)+r\,k^i+P_0^2\left (\frac{\partial k^i}{\partial x}\,F_y-\frac{\partial k^i}{\partial y}\,F_x\right )\ ,\end{equation}where 
\begin{equation}\label{87}
F=s^i\,k_i\ ,\end{equation}and the subscripts on $F$ denote partial differentiation with respect to $x$ and $y$. We note that $P_0$ and $k^i$ are given by (\ref{6}) and (\ref{8}). By (\ref{84}) and the 
useful formulas (\ref{33})--(\ref{35}) we easily find that $F$ is an $l=1$ spherical harmonic. Also we can rewrite (\ref{86}) in the general form of (\ref{26}) as
\begin{equation}\label{88}
X^i=w^i(u)+r\,K^i\ \ \ {\rm with}\ \ \ K^i=\left (\eta^{ij}+\frac{1}{r}\,s^{ij}\right )\,k_j\ ,\end{equation}showing that for large $r$ in Minkowskian space--time $K^i$ only differs from $k^i$ by an infinitesimal 
Lorentz transformation generated by the spin tensor. The Minkowskian line element is given by $ds_0^2=\eta_{ij}\,dX^i\,dX^j$ with $X^i$ given now by (\ref{86}). When $X^i=w^i(u)$ (no longer corresponding to $r=0$) 
is a geodesic we can choose $v^i=\delta ^i_4$ and the resulting line element is 
\begin{equation}\label{89}
ds_0^2=-(r^2+F^2)\,p_0^{-2}(dx^2+dy^2)+2\,d\Sigma\,\left\{dr+\frac{1}{2}(du-F_y\,dx+F_x\,dy)\right\}\ ,\end{equation}with $p_0$ given by (\ref{2}) and 
\begin{equation}\label{90}
d\Sigma=du-F_y\,dx+F_x\,dy\ .\end{equation}We note that the latter is \emph{not} an exact differential unless the spin vanishes. The Kerr solution with the three components of angular momentum 
$(m\,s^1, m\,s^2, m\,s^3)$ (since now $s^4=0$) is 
\begin{equation}\label{91}
ds^2=ds_0^2-\frac{2\,m\,r}{r^2+F^2}\,d\Sigma^2\ .\end{equation}If the world line $X^i=w^i(u)$ is \emph{not} a time--like geodesic then the Minkowskian line element is algebraically 
more complicated than (\ref{89}). For simplicity we shall henceforth neglect spin--spin terms and assume that the 4--acceleration is proportional to the spin on the basis that we expect 
geodesic motion if the spin vanishes. Now the Minkowskian part of the background line element, when the world--line $X^i=w^i(u)$ is not necessarily a time--like geodesic, is given by
\begin{eqnarray}
ds_0^2&=&-r^2\,P_0^{-2}\left\{dx^2+dy^2+\frac{2P_0^2\,F_y}{r^2}\,dx\,du-\frac{2P_0^2\,F_x}{r^2}\,dy\,du\right\}+2\,dr\,d\Sigma\nonumber\\
&&+2\,F_y\,du\,dx-2\,F_x\,du\,dy+(1-2\,h_0\,r)\,du^2\ .\label{92'}\end{eqnarray}With this preparation we write the line element of the general background space--time as
\begin{eqnarray}
ds^2&=&-r^2p^{-2}\Biggl\{(e^{\alpha}\,\cosh\beta\,dx+e^{-\alpha}\sinh\beta\,dy+a\,d\Sigma)^2\nonumber\\
&&+(e^{\alpha}\,\sinh\beta\,dx+e^{-\alpha}\cosh\beta\,dy+b\,d\Sigma)^2\Biggr\}+2\,dr\,d\Sigma+c\,d\Sigma^2\ ,
\nonumber\\
&&\label{92}\end{eqnarray}with
\begin{eqnarray}
p&=&P_0\,(1+q_2\,r^2+q_3\,r^3+\dots\ \ \ )\ ,\label{93}\\
\alpha&=&\alpha_1\,r+\alpha_2\,r^2+\dots\ \ \ \ ,\label{94}\\
\beta&=&\beta_1\,r+\beta_2\,r^2+\dots\ \ \ \ ,\label{95}\\
a&=&\frac{P_0^2\,F_y}{r^2}+\frac{a_{-1}}{r}+a_0+a_1\,r+\dots\ \ \ \ ,\label{96}\\
b&=&-\frac{P_0^2\,F_x}{r^2}+\frac{b_{-1}}{r}+b_0+b_1\,r+\dots\ \ \ \ ,\label{97}\\
c&=&c_0+c_1\,r+\dots\ \ \ \ .\label{98}\end{eqnarray}We now require this background line element to be a solution of Einstein's vacuum field equations
\begin{equation}\label{99}
R_{ab}=0\ .\end{equation}Equating to zero the powers of $r$ in the Ricci tensor components $R_{ab}$ we find the following expressions for the coefficients of the powers of $r$ in the 
expansions (\ref{93})--(\ref{98}):
\begin{equation}\label{100}
q_1=0=q_2\ ,\ \ \ \alpha_1=4\,F\beta_2\ ,\ \ \ \beta_1=-4\,F\alpha_2\ ,\end{equation}
\begin{eqnarray}
\alpha_2&=&\frac{1}{6}P_0^2\,R_{ijkl}(u)\,\frac{\partial k^i}{\partial x}\,k^j\,\frac{\partial k^k}{\partial x}\,k^l+O(F)\ ,\label{101}\\
\beta_2&=&\frac{1}{6}P_0^2\,R_{ijkl}(u)\,\frac{\partial k^i}{\partial x}\,k^j\,\frac{\partial k^k}{\partial y}\,k^l+O(F)\ .\label{102}\end{eqnarray}Also $a_{-1}=0=b_{-1}$ and
\begin{equation}\label{103}
a_0=-7P_0^2(\alpha_2F_y-\beta_2F_x)\ ,\ \ \ b_0=-7P_0^2(\alpha_2F_x+\beta_2F_y)\ ,\end{equation}
\begin{eqnarray}
a_1&=&\frac{2}{3}P_0^2\,R_{ijkl}(u)\,k^i\,v^j\,k^k\,\frac{\partial k^l}{\partial x}+O(F)\ ,\label{104}\\
b_1&=&\frac{2}{3}P_0^2\,R_{ijkl}(u)\,k^i\,v^j\,k^k\,\frac{\partial k^l}{\partial y}+O(F)\ ,\label{105}\end{eqnarray}and $c_0=1$ while
\begin{equation}\label{106}
c_1=-2\,h_0-FP_0^2\,\left (\frac{\partial}{\partial y}(P_0^{-2}a_1)-\frac{\partial}{\partial x}(P_0^{-2}b_1)\right )+5(a_1F_y-b_1F_x)\ .\end{equation}
The $O(F)$--terms not calculated here will get multiplied by $F$, as in (\ref{100}), (\ref{103}) and (\ref{106}) and will therefore not contribute to the end result since we are systematically 
neglecting spin--spin terms. In these formulas $R_{ijkl}(u)$ are the components of the Riemann curvature tensor in coordinates $X^i$ calculated on the world--line $X^i=w^i(u)$. As well 
as satisfying the algebraic symmetries of the Riemann tensor they satisfy Einstein's vacuum field equations $\eta^{ik}R_{ijkl}(u)=0$ on the world --line $X^i=w^i(u)$. 

We introduce the Kerr particle of small mass $m=O_1$ as a perturbation, which is predominantly the Kerr field for small values of $r$, of this background vacuum space--time. The 
simplest way to achieve this appears to be the expansions (remembering that we make no claim to uniqueness; see the comments following (\ref{70}) above): 
\begin{eqnarray}
d\hat s^2&=&-r^2\hat p^{-2}\Biggl\{(e^{\hat\alpha}\,\cosh\hat\beta\,dx+e^{-\hat\alpha}\sinh\hat\beta\,dy+\hat a\,d\Sigma)^2\nonumber\\
&&+(e^{\hat\alpha}\,\sinh\hat\beta\,dx+e^{-\hat\alpha}\cosh\hat\beta\,dy+\hat b\,d\Sigma)^2\Biggr\}+2\,dr\,d\Sigma+\hat c\,d\Sigma^2\ ,
\nonumber\\
&&\label{107}\end{eqnarray}with
\begin{eqnarray}
\hat p&=&\hat P_0\,(1+\hat q_2\,r^2+\hat q_3\,r^3+\dots\ \ \ )\ ,\label{108}\\
\hat\alpha&=&\hat\alpha_1\,r+\hat\alpha_2\,r^2+\dots\ \ \ \ ,\label{109}\\
\hat\beta&=&\hat\beta_1\,r+\hat\beta_2\,r^2+\dots\ \ \ \ ,\label{110}\\
\hat a&=&\frac{P_0^2\,F_y}{r^2}+\frac{\hat a_{-1}}{r}+\hat a_0+\hat a_1\,r+\dots\ \ \ \ ,\label{111}\\
\hat b&=&-\frac{P_0^2\,F_x}{r^2}+\frac{\hat b_{-1}}{r}+\hat b_0+\hat b_1\,r+\dots\ \ \ \ ,\label{112}\\
\hat c&=&-\frac{2\,m}{r}+\hat c_0+\hat c_1\,r+\dots\ \ \ \ .\label{113}\end{eqnarray}As always the hatted functions of $x, y, u$ differ from their background values by $O_1$--terms. In addition 
$\hat P_0$ again has the form (\ref{72}) involving the first order function $Q_1(x, y, u)$. Now we must impose the vacuum field equations to be satisfied by the metric given via the line element (\ref{107}) 
with sufficient accuracy to enable us to derive the equations of motion for the world line $X^i=w^i(u)$ in the background space--time, in first approximation. The vacuum field equations provide 
us with a set of equations which parallel (\ref{76})--(\ref{78}) in the Reissner--Nordstr\"om example. In addition to $\hat q_1=O_2,  \hat q_2=O_2$ we find that
\begin{eqnarray}
\hat a_0&=&-7P_0^2(\alpha_2F_y-\beta_2F_x)-2\,m\,a_1+O(mF)+O_2\ ,\label{114}\\
\hat b_0&=&-7P_0^2(\alpha_2F_x+\beta_2F_y)-2\,m\,b_1+O(mF)+O_2\ ,\label{115}\end{eqnarray}
\begin{equation}\label{116}
\hat a_{-1}=-4mF\,b_1+O_2\ ,\ \ \ \ \hat b_{-1}=4mF\,a_1+O_2\ ,\end{equation}and
\begin{eqnarray}
\hat c_0&=&1+\Delta Q_1+2\,Q_1-10mF\,P_0^2\left (\frac{\partial}{\partial y}(P_0^{-2}a_1)-\frac{\partial}{\partial x}(P_0^{-2}b_1)\right )\nonumber\\
&&-8m\,(a_1F_y-b_1F_x)+O_2\ .\label{117}\end{eqnarray}In the ramaining field equation, which yields the analogue of (\ref{76}) and thence the analogue of (\ref{77}), the functions 
$\hat a_0$ and $\hat b_0$ are multiplied by $m$ and so the uncalculated terms $O(mF)$ in (\ref{114}) and (\ref{115}) will not contribute to the equations of motion in first approximation 
(i.e. neglecting $O_2$--terms). The equation we obtain, which is the analogue of (\ref{77}), is 
\begin{equation}\label{118}
(\Delta +2)(Q_1+2m\,J_1+m\,J_2)=6m\,a_i\,k^i-6m\,R_{ijkl}(u)\,k^i\,v^j\,s^{kl}+A(u)+O_2\ ,\end{equation}where $A(u)=O_1$ is a function of integration as in (\ref{77}) and $J_1$ and $J_2$ are 
spin--curvature terms given by
\begin{equation}\label{119}
J_1=P_0^2F\,R_{ijkl}(u)\,k^i\,v^j\,\frac{\partial k^k}{\partial y}\frac{\partial k^l}{\partial x}+\frac{1}{5}R_{ijkl}(u)\,k^i\,v^j\,s^{kl}\ ,\end{equation} which is an $l=3$ spherical harmonic and
\begin{equation}\label{120}
J_2=P_0^2\,R_{ijkl}(u)\,s_m\,v^i\,k^j\,v^k\,\left (\frac{\partial k^l}{\partial x}\frac{\partial k^m}{\partial y}-\frac{\partial k^l}{\partial y}\frac{\partial k^m}{\partial x}\right )\ ,\end{equation}which is an 
$l=2$ spherical harmonic, results 
which are obtained using the useful formulas (\ref{33})--(\ref{35}). The first two terms on the right hand side of (\ref{118}) are $l=1$ spherical harmonics and so, just as in the case of (\ref{77}), the solution $Q_1$ will be free 
of directional singularities provided the sum of these terms is zero or at most small of second order, i.e.
\begin{equation}\label{121}
m\,a_i\,k^i=m\,R_{ijkl}(u)\,k^i\,v^j\,s^{kl}+O_2\ ,\end{equation}for all possible $k^i$ and thus we arrive at the equations of motion of the Kerr particle in first approximation:
\begin{equation}\label{122}
m\,a_i=m\,R_{ijkl}(u)\,v^j\,s^{kl}+O_2\ .\end{equation}The algebraic form of the right hand side of these equations is, perhaps, what one would expect. The numerical factor of unity distinguishes 
these equations of motion from those of Mathisson--Papapetrou for a spinning \emph{test} particle. We have here a model of a Kerr particle and it is manifestly \emph{not} a test particle. The rest mass $m$ is 
already small and making it smaller does not make it a test particle since so long as $m\neq0$ its presence perturbs the background space--time (\ref{92}). However in the present context 
we can say something about spinning test particles and this is the topic we turn to now.

\section{Spinning Test Particle}
Since we are now concerned with test particles moving in vacuum gravitational fields we will concentrate on what we called background space--times above. We first consider the non--spinning 
test particle whose world line, according the geodesic hypothesis, is a time--like geodesic. With the geodesic hypothesis accepted for a non--spinning test particle it is relatively easy to devise 
a strategy, using the formalism of this paper, which will lead us to the equations of motion of a spinning test particle. As in Section 3 we will neglect spin--spin terms, but this restriction can be relaxed 
(and indeed has been by Florian Bolgar in his Internship Report to the \'Ecole Normale Sup\'erieure (2012)). The background vacuum space--time with the spin vector $s^i=0$ has line element given by (\ref{92})--(\ref{98}) 
with $F=0$ and the coefficients of the powers of $r$ are given by (\ref{100})--(\ref{106}) with $F=0$. Thus in coordinates $x^i=(x, y, r, u)$, with $i=1, 2, 3, 4$, the only coordinate component of 
the metric tensor involving the 4--acceleration $a^i$ of the world line $X^i=w^i(u)$ is
\begin{equation}\label{123}
g_{44}=1-2r\,h_0+O(r^2)\ .\end{equation}The coefficient of $-2r$ on the right hand side here is the $l=1$ spherical harmonic $h_0$, which vanishes if and only if $a^i=0$. In this case the 
world line $X^i=w^i(u)$ is a time--like geodesic and is thus the history of a test particle moving in the vacuum gravitational field modeled by the space--time with line element (\ref{92}) with $F=0$. Now 
we consider a particle with spin $s^i\neq0$ having world line $X^i=w^i(u)$ in the space--time with line element (\ref{92}) with $F\neq0$. Now (\ref{123}) is replaced by
\begin{equation}\label{124}
g_{44}=1-2r\,\left (h_0-\frac{1}{2}R_{ijkl}(u)\,k^i\,v^j\,s^{kl}+J_2\right )+O(r^2)\ ,\end{equation}with $J_2$ given by (\ref{120}). The first two terms in the coefficient of $-2r$ here are $l=1$ spherical harmonics while 
the third term $J_2$ is an $l=2$ spherical harmonic. If the equations of motion of a spinning test particle are obtained in the same way as those of a non--spinning test particle then equating to zero the 
$l=1$ terms in the coefficient of $-2r$ in (\ref{124}) results in 
\begin{equation}\label{125}
a_i\,k^i=h_0=\frac{1}{2}R_{ijkl}(u)\,k^i\,v^j\,s^{kl}\ ,\end{equation}for all possible values of $k^i$ and thus the equations of motion of a spinning test particle, neglecting spin--spin terms, are
\begin{equation}\label{126}
a_i= \frac{1}{2}R_{ijkl}(u)\,v^j\,s^{kl}\ ,\end{equation}in agreement with Mathisson \cite{Math} and Papapetrou \cite{Papa}. 

\section{Discussion}
Perhaps the most striking aspect of the technique presented here for extracting equations of motion from field equations is the absence of infinities arising in the calculations, either 
in the form of infinite self energy of the Dirac type \cite{DeB} or in the form of divergent integrals. We have already indicated following (\ref{81}) above that in the case of the DeWitt and Brehme \cite{DeB} 
work this may be due to the fact that they have considered a different problem to the one we have described in Section 2. Much work in recent years has been done under the general heading 
of ``the self force problem" and, fortunately, this work is well represented in the current volume, offering the interested student a direct comparison with our work and the challenge of 
connecting the different points of view.

\end{document}